\pdfoutput=1

\documentclass[11pt]{article}

\usepackage{ACL2023}

\usepackage{times}
\usepackage{latexsym}

\usepackage[T1]{fontenc}

\usepackage[utf8]{inputenc}

\usepackage{microtype}

\usepackage{inconsolata}

\usepackage{graphicx}

\title{An LLM-based Knowledge Synthesis and Scientific Reasoning Framework for Biomedical Discovery}

\author{
  \textbf{Oskar Wysocki\textsuperscript{1,2}},
  \textbf{Magdalena Wysocka\textsuperscript{2}},
  \textbf{Danilo S. Carvalho\textsuperscript{2}},
  \textbf{Alex Bogatu\textsuperscript{2}},
\\
  \textbf{Danilo Gusicuma\textsuperscript{1}},
  \textbf{Maxime Delmas\textsuperscript{1}},
  \textbf{Harriet Unsworth\textsuperscript{2}},
  \textbf{Andr\'e Freitas\textsuperscript{1,2,3}}
\\
\\
  \textsuperscript{1}Idiap Research Institute, Switzerland\\
  \textsuperscript{2}National Biomarker Centre, CRUK-MI, Univ. of Manchester, United Kingdom\\
  \textsuperscript{3}Department of Computer Science, Univ. of Manchester, United Kingdom
\\
  \small{
    \textbf{Correspondence:} {firstname.lastname}@idiap.ch\textsuperscript{1}{firstname.lastname}@manchester.ac.uk\textsuperscript{2}
  }
}

\begin{document}
{\makeatletter\acl@finalcopytrue
  \maketitle
}
\begin{abstract}
We present BioLunar, developed using the Lunar framework, as a tool for supporting biological analyses, with a particular emphasis on molecular-level evidence enrichment for biomarker discovery in oncology. The platform integrates Large Language Models (LLMs) to facilitate complex scientific reasoning across distributed evidence spaces, enhancing the capability for harmonizing and reasoning over heterogeneous data sources. Demonstrating its utility in cancer research, BioLunar leverages modular design, reusable data access and data analysis components, and a low-code user interface, enabling researchers of all programming levels to construct LLM-enabled scientific workflows. By facilitating automatic scientific discovery and inference from heterogeneous evidence, BioLunar exemplifies the potential of the integration between LLMs, specialised databases and biomedical tools to support expert-level knowledge synthesis and discovery.

\end{abstract}

\section{Introduction}

Contemporary biomedical discovery represents a prototypical instance of complex scientific reasoning, which requires the coordination of controlled \textit{in-vivo}/\textit{in-silico} interventions, complex multi-step data analysis pipelines and the interpretation of the results under the light of previous evidence (available in different curated databases and in the literature) \cite{Paananen2019, NICHOLSON20201414}. This intricacy emerges out of the inherent complexity of biological mechanisms underlying organism responses, which are defined by a network of multi-scale inter-dependencies \cite{BOGDAN}. While more granular data is being generated by the evolution of instruments, assays and methods, and the parallel abundance of experimental interventions \cite{DRYDEN}, there a practical barrier for integrating and cohering this evidence space into a specific context of analysis.

Within biomedical discovery, the language interpretation capabilities of Large Language Models (LLMs) can provide an integrative framework for harmonising and reasoning over distributed evidence spaces and tools, systematising and lowering the barriers to access and reason over multiple structured databases, textual bases such as PubMed, enriching the background knowledge through specialised ontologies and serving as interfaces to external analytical tools (e.g. mechanistic/perturbation models, gene enrichment models, etc). In this context, LLMs can serve as a linguistic analytical layer which can reduce the syntactic impedance across diverse functional components: once an adapter to an external component is built it can be integrated and reused in different contexts, creating a monotonic increase of functional components. Complementarily, from a Biomedical-NLP perspective, in order to address real-world problems, LLMs need to be complemented with mechanisms which can deliver contextual control (e.g. via Retrieval Augmented Generation: RAG: access the relevant background knowledge and facts) and perform the analytical tasks which are integral to contemporary biomedical inference ('toolforming'). 

Emerging LLM-focused coordination frameworks such as LangChain\footnote{\url{https://python.langchain.com}}, Flowise\footnote{\url{https://github.com/FlowiseAI/Flowise}} and Lunar\footnote{\url{https://lunarbase.ai}} provide the capabilities to deliver a composition of functional components, some of them under a low-code/no-code use environment, using the abstraction of workflows. While there are general-purpose coordination frameworks, there is a lack of specialised components for addressing biomedical analyses. 

In this paper we demonstrate BioLunar, a suite of components developed over the Lunar environment to support biological analyses. We demonstrate the key functionalities of the platform contextualised within a real-use case in the context of molecular-level evidence enrichment for biomarker discovery in oncology.



\section{BioLunar}

\begin{figure*}[htbp]
\centering
\includegraphics[width= .99\textwidth]{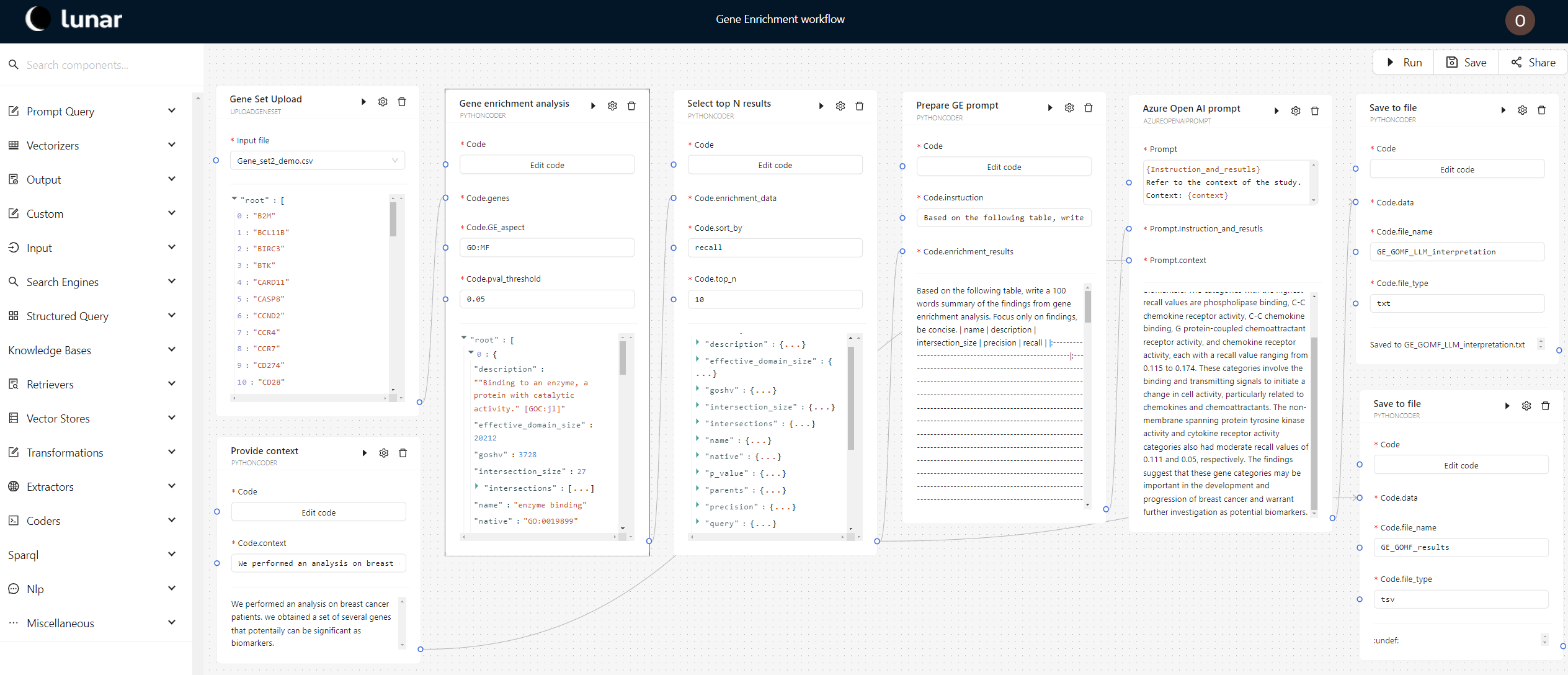}
\caption{BioLunar interface. An exemplary workflow of Gene Enrichment with an input gene set, knowledge base query and LLM interpretation components. }
\label{fig:Lunar_screen_GE_workflow}
\end{figure*}

BioLunar enables the creation of LLM-based biomedical scientific workflows using software components with standardised APIs. A workflow is composed of components and subworkflows connected through input-output relationships, and are capable of handling multiple inputs. In the user interface, components are clustered according to their function (see Fig.\ref{fig:Lunar_screen_GE_workflow}). Creating a workflow does not require programming knowledge since components are predefined and merely require data inputs or parameter settings. However, for users who wish to write custom code, 'Python Coder' and 'R Coder' components are provided, enabling the definition of custom methods. These custom components can be saved and subsequently accessed in the 'Custom' group tab.

In the paper we describe an exemplar \textit{biomedical workflow} designed to integrate evidence and infer conclusions from bioinformatics pipeline results. Specifically, the \textit{biomedical workflow} queries expert knowledge bases (KBs) that continuously compile clinical, experimental, and population genetic study outcomes, aligning them with assertions relevant to the significance of the observed gene or variant. It then employs Natural Language Inference (NLI) (via LLM) to integrate and harmonise the evidence space and interpreting the results, culminating in a comprehensive summary for the entire gene set input. This interpretation takes into account the bioanalytical context supplied by the user.


\subsection{Exemplar Workflow}

Next-generation sequencing (NGS) assays play a pivotal role in the precise characterisation of tumours and patients in experimental cancer treatments. NGS findings are essential to guide the design of novel biomarkers and cancer treatments. Nevertheless, the clinical elucidation of NGS findings subsequent to initial bioinformatics analysis often requires time-consuming manual analysis procedures which are vulnerable to errors. The interpretation of molecular signatures that are typically yielded by genome-scale experiments are often supported by pathway-centric approaches through which mechanistic insights can be gained by pointing at a set of biological processes. Moreover, gene and variant enrichment benefits from heterogeneous curated data sources which pose challenges to seamless integration. Furthermore, there are different levels of supporting evidence and therefore prioritising conclusions is crucial. Automating evidence interpretation, knowledge synthesis and leveraging evidence-rich gene set reports are fundamental for addressing the challenges in precision oncology and the discovery of new biomarkers.









\subsection{User interface}

The user interface facilitates an agile workflow construction by enabling users to select and arrange components via drag-and-drop from functionally grouped categories, such as, i.a.: 'Prompt Query' featuring NLI components, 'Knowledge Bases' components, 'Extractors' for retrieving files from zip archives or extracting text and tables from PDF files, and 'Coders', which allow for the creation of custom components using Python or R scripts.

Components allow for individual execution, edition, or configuration adjustment via a visual interface. Workflows can be executed, saved, or shared. Each component has designated input and output capabilities, enabling seamless integration where the output from one can directly feed into another. Users have the flexibility to manually input values if no direct connection is established. Additionally, a component's output can feed into multiple components. The system's architecture supports effortless expansion, adding branches and components without affecting the existing workflow, thus facilitating scalable customization to meet changing requirements.
The user interface with an example of a workflow is presented in Fig.\ref{fig:Lunar_screen_GE_workflow} and in demo video \url{https://youtu.be/Hc6pAA_5Xu8}.

\subsection{Knowledge bases}


The current framework integrates a diverse set of knowledge bases which are relevant for precision oncology.
To identify gene mutations as biomarkers for cancer diagnosis, prognosis, and drug response, we integrated CIViC\footnote{\url{https://civicdb.org}} and OncoKB\footnote{\url{https://www.oncokb.org}}. CIViC provides molecular profiles (MPs) of genes, each linked to clinical evidence, with a molecular score indicating evidence quality, assessed by human annotators. The Gene Ontology\footnote{\url{https://geneontology.org}} (GO) offered gene function insights, and the Human Protein Atlas\footnote{\url{https://www.proteinatlas.org}} supplied a list of potential drug targets and transcription factors. We employed COSMIC\footnote{\url{https://cancer.sanger.ac.uk/cosmic}} for somatic mutation impacts in cancer, the largest resource in this field. Our analysis also included KEGG\footnote{\url{https://www.kegg.jp/kegg/}}, Reactome\footnote{\url{https://reactome.org}}, and WikiPathways\footnote{\url{https://www.wikipathways.org}} for pathway information, enriching our investigation with scientific literature via PubMed's API \footnote{\url{https://pubmed.ncbi.nlm.nih.gov}}.

In the following subsections, we showcase examples of components, subworkflows, and workflows constructed using the BioLunar framework, motivated by the biomarker discovery/precision oncology themes.

\subsection{Construction and reuse of specialised prompts}

BioLunar employs standard LLM interfaces, allowing the use of different models according to users' preferences. The prompt components allows for the composition of specialised prompt chains which can be later reused, defining a pragmatic pathway for specialised Natural Language Inference (NLI) via prompt decomposition/composition. This approach allows for the creation of reasoning chains that combines user's instructions with the results of database queries and analyses from specialised tools within the context of the study. An instantiated example of the \textit{Azure Open AI prompt} is described in Fig.\ref{fig:Lunar_screen_GE_workflow}.

\subsection{Subworkflow component}

\begin{figure*}[tb]
\centering
\includegraphics[width= .7\textwidth]{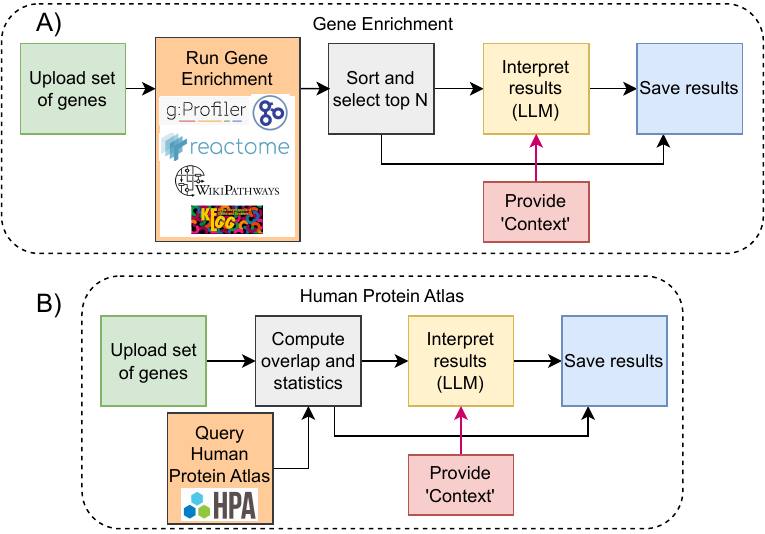}
\caption{A) Gene Enrichment workflow - uses the \textit{gprofiler} API to access i.a. Gene Ontology, KEGG, WikiPathways, Reactome; B) Human Protein Atlas workflow. Compares and interprets the input and reference gene sets.}
\label{fig:diagram_GE_HPA}
\end{figure*}

\begin{figure*}[tb]
\centering
\includegraphics[width= .98\textwidth]{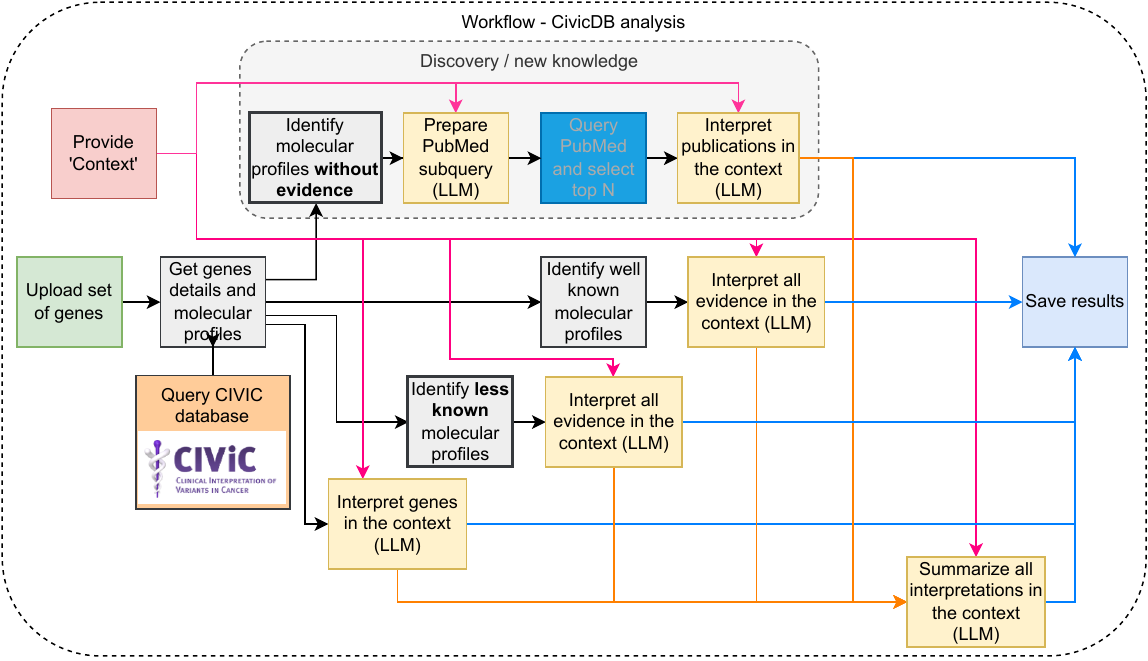}
\caption{CIVIC evidence analysis workflow. prompt-based NLI components are fed by both the results and context of the analysis in order to produce relevant evidence-based conclusions.}
\label{fig:diagram_CIVIC_workflow}
\end{figure*}

The \textit{subworkflow} component enables the reuse of an existing workflow within another workflow, functioning as a component with specified inputs and outputs. This feature simplifies the composition of more complex workflows and avoids the repetition of defining identical steps for the same task. Subworkflows can be selected like other components from the left panel in the interface, offering access to all available workflows for easy integration. Examples of subworkflows are presented in Fig.\ref{fig:diagram_GE_HPA},\ref{fig:diagram_CIVIC_workflow}.

\subsection{Gene Enrichment subworkflow}

One example of a specialised subworkflow is the \textit{Gene Enrichment subworkflow} (Fig.\ref{fig:Lunar_screen_GE_workflow},\ref{fig:diagram_GE_HPA}A) begins with uploading the targeted gene sets. Then a component accesses a specific KB — such as Gene Ontology, KEGG, Reactome, or WikiPathways—defined by the user, using \textit{gprofiler} API\footnote{\url{https://biit.cs.ut.ee/gprofiler/page/apis}}. This component identifies gene groups with a statistically significant overlap with the input gene set, according to a Fisher's test, and calculates p-values, recall, and precision. The user then specifies a variable to rank these groups and selects the top N for further analysis. The output includes both a interpretation performed by an NLI component (through LLM) and a table featuring the names, descriptions, and statistics of the top N selected groups. 

\subsection{Human Protein Atlas subworkflow}

In the \textit{Human Protein Atlas subworkflow}, given a gene set, an associated external KB is queried by selecting `Transcription factors' from the HPA database using a dedicated query-database connector. A reusable component, 'Analyze overlap', then identifies genes that overlap and calculates relevant statistics. Similarly to the \textit{Gene Enrichment subworkflow}, the results are interpreted by an prompt-based NLI component and presented alongside a table summarising the findings (Fig.\ref{fig:diagram_GE_HPA}B,\ref{fig:Lunar_screen_HPA_workflow}).

\subsection{CIVIC subworkflow}

This \textit{subworkflow} exemplifies a more complex composition of components (Fig.\ref{fig:diagram_CIVIC_workflow}). This subworkflow initiates by querying the CIVIC database for input genes, yielding, among other things, gene descriptions in clinical contexts, and their variants and molecular profiles (MPs), which are essential for the final interpretation. Additionally, users specify the analysis context, including aspects such as cancer types or subtypes, treatments, populations, etc. Initially, gene descriptions are analysed by a prompt-based NLI component within this defined context. Subsequently, MPs scored below a predefined threshold (set at a MP score of 10) are tagged as \textit{less known}, reflecting lower scientific evidence and ranking by CIVIC annotators. The evidence supporting these lesser-known MPs is then interpreted by a prompt-based NLI component, considering the broader analysis context. Conversely, evidence from \textit{well-known} MPs, scoring above 10, undergoes a similar interpretation process. 

For genes without identified MPs in CIVIC, a sequence of components perform further evidence retrieval from PubMed. An NLI module generates context-based keywords for PubMed queries, which are combined with the names of genes lacking MPs. A 'PubMed search' component then retrieves $N$ publications, including metadata, citation counts and MeSH terms (used later for context alignment validation). The abstracts of these publications are interpreted by an NLI module in the context of the analysis. 
    
All clinical evidence interpretations are then succinctly summarised by via a prompt component, taking into account the context of the analysis. These interpretations, along with tabular results, constitute the output.

\subsection{Bioworkflow - comprehensive analysis for a set of genes.}
\label{bioworkflow}
\begin{figure*}[htbp]
\centering
\includegraphics[width= .8\textwidth]{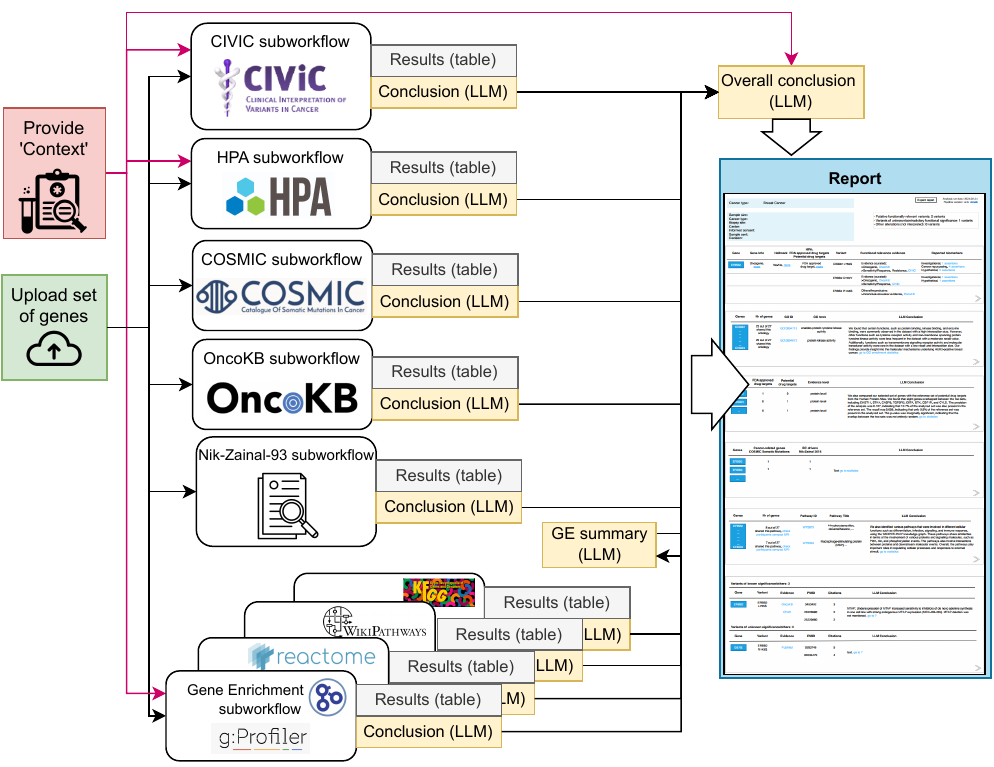}
\caption{Diagram of the Bioworkflow. }
\label{fig:diagram_bioworkflow}
\end{figure*}

The exemplar \textit{bioworkflow} composes multiple subworkflows (Fig.\ref{fig:diagram_bioworkflow}), each dedicated to a specific multi-step and specialised task, which are typically defined by the composition of heterogeneous components, most commonly connectors and query instance components to specialised databases (e.g. CIVIC, HPA, PubMed, OncoKB), external specialised analytical tools (toolformers for gene enrichment analysis) and chains of specialised interpretation prompts (e.g. selection, filtering, extraction, summarisation). This setup forms a comprehensive workflow which exemplifies the close dialogue between LLMs and genomic analysis, encompassing gene enrichment, comparison with reference gene sets, and access to evidence within an experimental medicine setting. Additionally, it queries PubMed publications within the CIVIC component to seek evidence for molecular profiles not yet described. Its componentised architecture facilitates the extensibility of the workflow with new sources, prompts and external tools. Conclusions drawn from each subworkflow are interpreted within the analysis context, being integrated in a comprehensive summary. All findings are compiled in a report, exported as a PDF file.

\subsection{Software description}

BioLunar uses the \textit{LunarVerse} backend for its operations. LunarVerse is downloaded and installed by the setup script included with the demonstration code. Some of its components need user specific configuration to work, such as private API keys, which are defined in a configuration file indicated in the setup instructions. LunarVerse is distributed under a \textit{open software} license. The workflow can also be operated via a graphical interface (\textit{LunarFlow}) 

Running a workflow can be done in two ways: i) directly, by calling the LunarVerse engine on a specified workflow descriptor file; ii) through the Web interface, by pressing the ``Run'' button.

The first way is the default one in the demonstration code. It returns a copy of the workflow descriptor, with all component output fields filled, which is then used to extract and filter the desired outputs, based on the component labels. It is also the best way to automate multiple workflow runs and to integrate their outputs into other systems.The supporting code is available at \url{https://github.com/neuro-symbolic-ai/lunar-bioverse-demo}.

\subsection{Report}

\begin{figure*}[t]
\centering
\includegraphics[width= .9\textwidth]{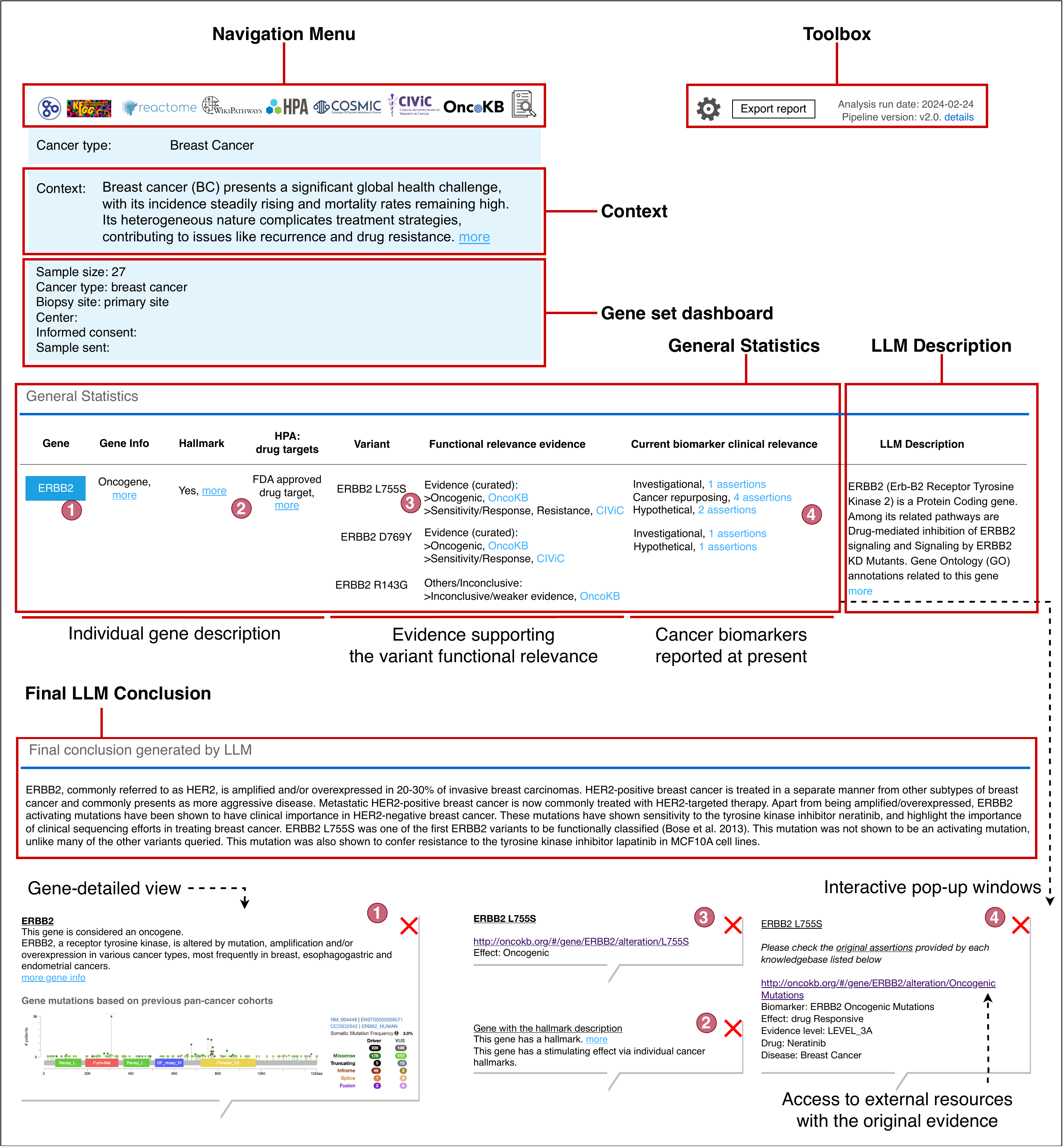}
\caption{The BioLunar report's overview, produced by \textit{Bioworkflow}.} 
\label{fig:report}
\end{figure*}

The \textit{Bioworkflow}, as outlined in point \ref{bioworkflow}, generates a report in PDF (Fig.\ref{fig:report}) format that begins by outlining the context of the study, analysis details, dates, and software versions at the top. The report is enhanced with hyperlinks for easy navigation to specific sections.

A "General Statistics" table provides a comprehensive overview of key metrics aggregated from all components, aiming to consolidate information for each gene throughout the analysis, with hyperlinks directing to the report sections where this information originates.

Subsequent sections categorise genes into various tables based on biological aspects and the KBs consulted. These include Molecular Function for genes sharing ontologies, drug target checks based on the Human Protein Atlas, assessments of cancer-related genes, Pathway Analysis and Mapping via WikiPathways, and classification of gene alterations by clinical relevance. By correlating genes with known functional information, the workflow identifies statistically significant enriched terms and summarizes these findings using LLM, which also furnishes evidence.

LLM interprets each table, offering textual conclusions relevant to the analysis context. A final summary, crafted using LLM, synthesizes all results within the given context.
Importantly, all LLM interpretations are grounded in concrete evidence, with sources cited alongside the narrative. This approach underscores the rigor of the analysis by highlighting distinct sources that substantiate the relevance of each gene and variant.

\section{Case study}

To demonstrate the capabilities of the \textit{Bioworkflow}, we analyzed outputs in two different scenarios, each producing a distinct set of genes from separate bioinformatics analyses. We entered these gene sets along with their analysis contexts into the \textit{Bioworkflow} and executed it. Subsequently, we qualitatively assessed the output reports (see Fig.\ref{fig:scenario1_report},\ref{fig:scenario2_report}), considering both the statistical data and the interpretations provided by the prompt-based NLI modules. 

In Scenario 1, the user aims to explore the unique molecular characteristics of HER2-low breast cancer to determine if it constitutes a distinct category within breast cancer types, where the  input genes are ERBB2, ESR1, PIK3CA, CBFB, SF3B. The report shows genomic alterations and genomic signatures that were identified, including ERBB2 amplification, mutations in PIK3CA and ESR1, which are important biomarkers in the selection of breast cancer treatment. For the remaining two genes, evidence was found confirming that these are new, significantly mutated genes for which there is pre-clinical evidence of actionability in clinical practice.


In Scenario 2, the user aims to discover new genes that could lead to more accurate breast cancer diagnoses, enhancing treatment strategies and addressing the disease's complexity. His numerical analysis resulted in a set of genes (DIXDC1, DUSP6, PDK4, CXCL12, IRF7, ITGA7, NEK2, NR3C1) that require investigation. The report informs that none of the genes is an oncogene (confirmation according to OncoKB), two of the genes are potential drug targets and one is FDA approved drug targets. According to the KEGG-based enrichment analysis, these genes were mainly enriched through several signaling pathways including tumor necrosis factor (TNF) signaling pathway. Using LLMs in conjunction with a PubMed search component, papers were searched in PubMed that describe various gene variants and the genes have been indicated as prospective biomarkers associated with breast cancer.

Note that in scenario 2, for genes lacking molecular profiles in the KB, a search in PubMed was conducted. This approach enables the workflow to automatically uncover and search for non-obvious and previously unknown relationships. Essentially, if a gene is absent from the database, it suggests that its relevance is relatively novel and not yet documented. Therefore, seeking out the most recent publications that describe this gene within the analysis context represents a significant advantage, provided by the workflow that integrates various components.


 



\section{Related Work}

\textbf{Bioinformatics Pipelines}
Over the past decade, three scientific workflow management systems such as Galaxy \cite{galaxy2022galaxy}, Snakemake \cite{koster2012snakemake}, and Nextflow \cite{di2017nextflow}, have been instrumental to bioinformaticians to systematise their complex analytical processes. Nextflow targets bioinformaticians and facilitates gene enrichment analysis, annotate biological sequences, and perform gene expression analysis by including modules supported by various bioinformatics tools. These workflow systems are currently centred around the composition of specialised bioinformatics software, configuration parameters and supporting datasets, facilitating reuse and reproducibility. In contrast, this paper explores the concept on using LLMs within a specialised workflow environment to support the interpretation and integration of multiple analytical processes.

\section{Conclusion}
In this paper we provided a demonstration of a scientific workflow based on LLMs to support specialised gene analyses using oncology and gene enrichment as a driving motivational scenario. The framework is built using the Lunar framework and allows for the composition of specialised analytical workflows, integrating external databases (Retrieval Augmented Generation), external tools (ToolFormers) and contextualised chains of LLM-based interpretation. The paper highlights that a workflow environment with specialised components for RAG, ToolFormers and a set of specialised prompts-based Natural Language Inference can serve as the foundation for streamlining and automating complex analytical process within a biomedical setting. . We showcase analytical applications within the biomedical domain, particularly in oncology, constructively progressing towards more complex gene analysis workflows. The developed \textit{bioworkflow} demonstrates the LLMs can be instrumental in enabling a complex end-to-end highly-specialised analytical workflow, in a reproducible manner, supporting the integration of heterogeneous evidence,  synthesising conclusions and while simultaneously documenting and linking to the data sources within a comprehensive output report. The proposed workflow is based on a low-code paradigm that enables domain experts, regardless of their programming skills, to construct and scientific workflows enabled by generaqtive AI amethods.

\section*{Limitations}
\begin{itemize}
    \item The current demonstration uses external LLM-based APIs but can be adapted to open source LLM models.
    \item The LLM-based inferences require a critical supporting quantitative evaluation and hallucinations are possible. The current workflow is motivated by a hypothesis generation process, which is fully human supervised and does not have direct clinical applications.
\end{itemize}




\section*{Acknowledgements}
This work was partially funded by The Ark foundation, by the European Union’s Horizon 2020 research and innovation program (grant no. 965397) through the Cancer Core Europe DART project, and by the Manchester Experimental Cancer Medicine Centre and the NIHR Manchester Biomedical Research Centre.

\bibliography{anthology,custom}
\bibliographystyle{acl_natbib}

\newpage
\appendix

\section{Appendix}
\label{sec:appendix}

\counterwithin{figure}{section}

\begin{figure*}[htbp]
\centering
\includegraphics[width= .7\textwidth]{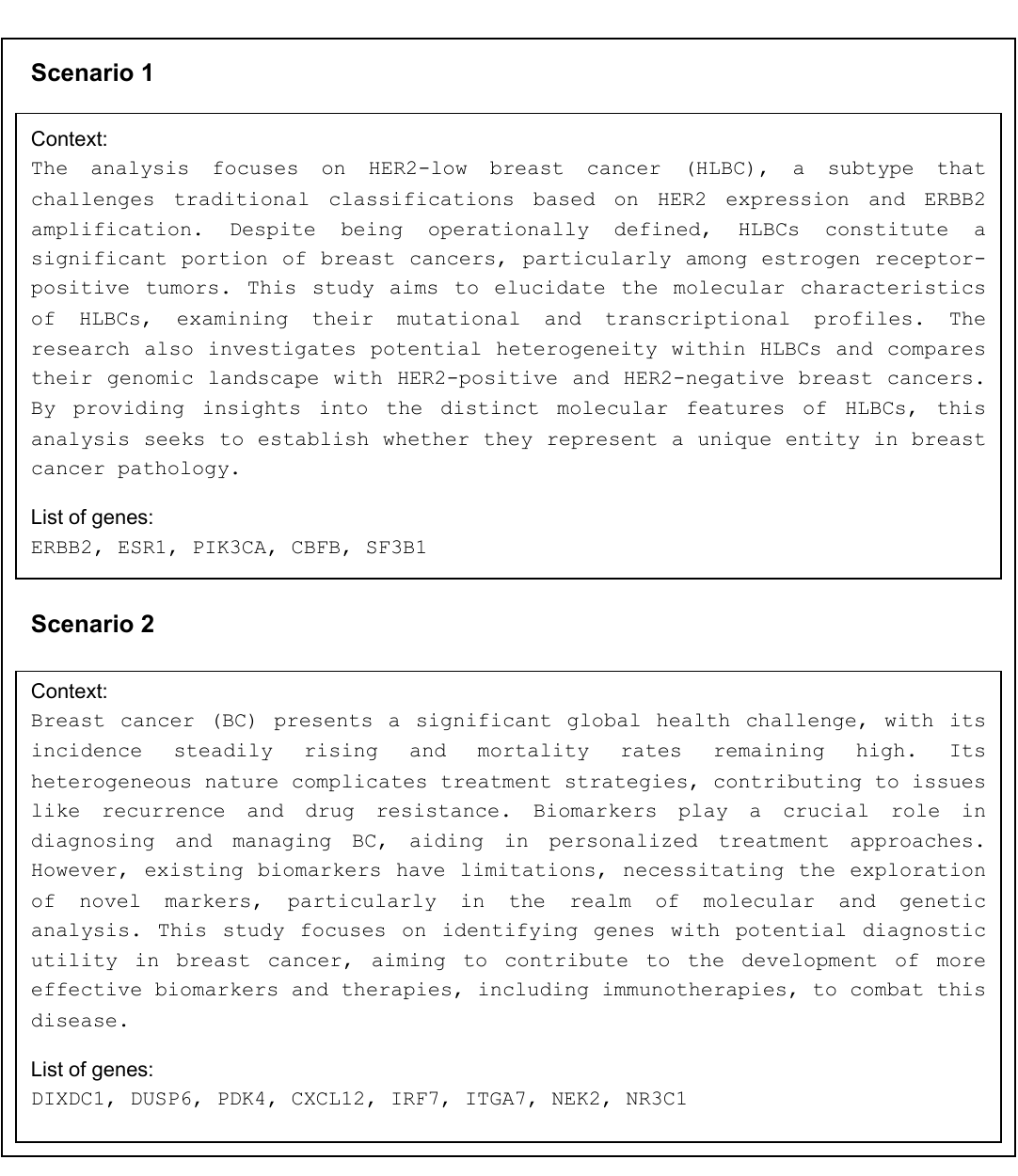}
\caption{User-defined context of the analysis, including aspects like cancer types or subtypes, treatments, populations, for Scenario 1 and 2.}
\label{fig:context}
\end{figure*}

\begin{figure*}[htbp]
\centering
\includegraphics[width= .99\textwidth]{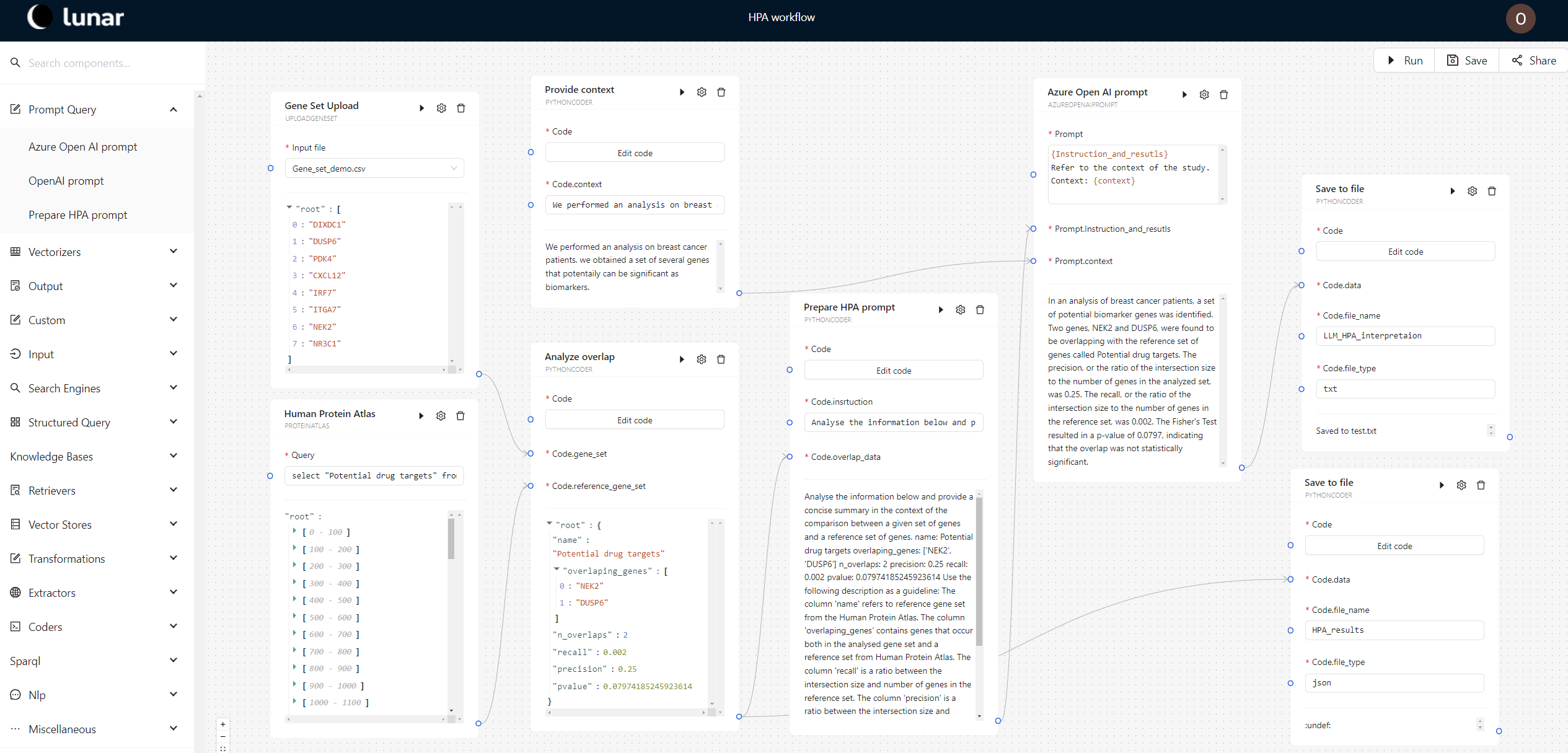}
\caption{Human Protein Atlas workflow in the BioLunar interface.}
\label{fig:Lunar_screen_HPA_workflow}
\end{figure*}

\begin{figure}[htbp]
\centering
\includegraphics[width= .5\textwidth]{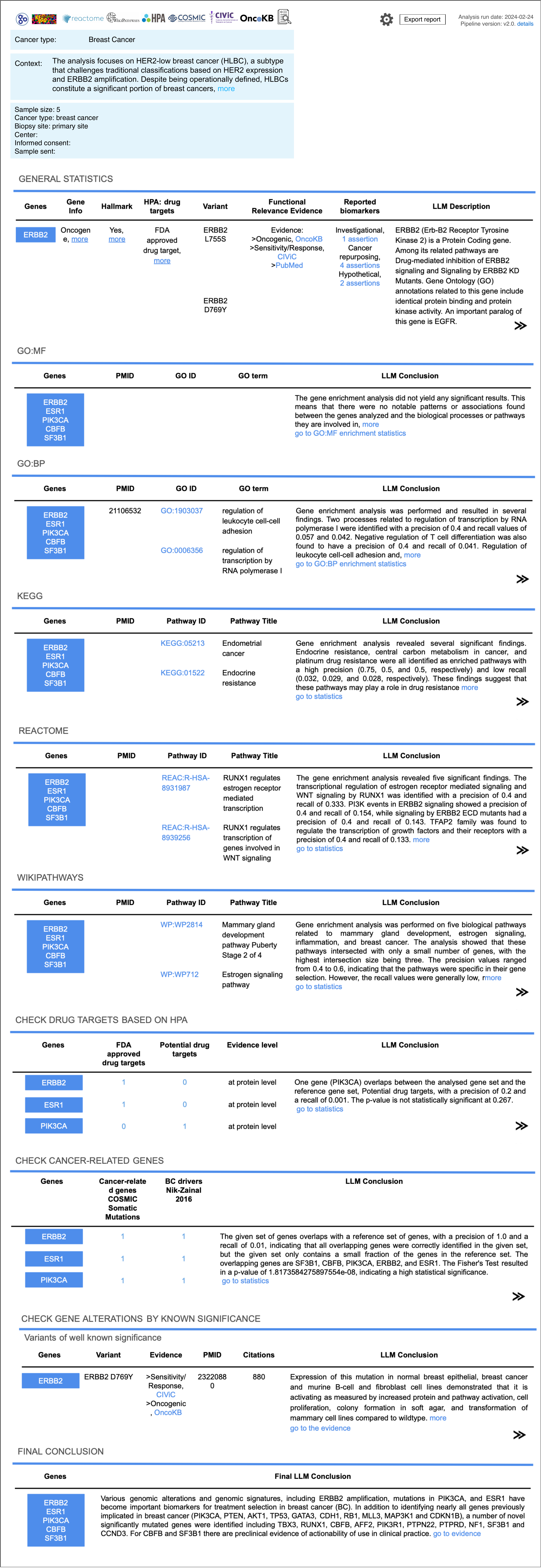}
\caption{The BioLunar report, produced by \textit{Bioworkflow} for Scenario 1} 
\label{fig:scenario1_report}
\end{figure}

\begin{figure}[htbp]
\centering
\includegraphics[width= .48\textwidth]{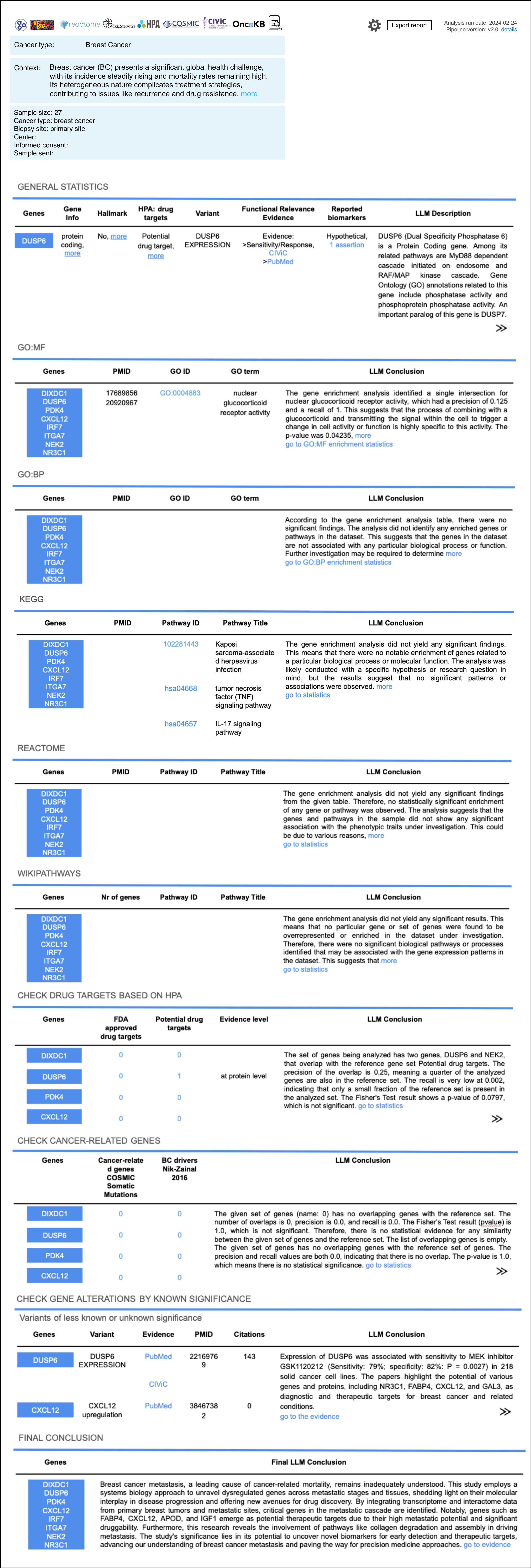}
\caption{The BioLunar report, produced by \textit{Bioworkflow} for Scenario 2.}
\label{fig:scenario2_report}
\end{figure}

\end{document}